\documentclass[twocolumn,showpacs,preprintnumbers,amsmath,amssymb]{revtex4}

\usepackage{graphicx}
\usepackage{dcolumn}
\usepackage{bm}

\begin{document}

\title{Spontaneous Emission of Charged Bosons from Supercritical Point Charges}

\author{Sang Pyo Kim}
\affiliation{Department of Physics, Kunsan National University, Kunsan 573-701, Korea}\email{sangkim@kunsan.ac.kr}
\medskip

\date{\today}

\begin{abstract}
We study the spontaneous emission of charged bosons from supercritical Coulomb potentials and charged black holes.
We find the exact emission rate from the Bogoliubov transformation by applying the tunneling boundary condition on the Jost functions
at the asymptotic boundaries. The emission rate for charged bosons in the supercritical Coulomb potential
increases as the charge $Z\alpha > 1/2$ of the superatom and the energy of the bosons increase
but is suppressed for large angular momenta. We discuss physical implications of the emission of charged bosons from superatoms and charged black holes.
\end{abstract}

\maketitle

\section{Introduction}\label{introduction}

The spontaneous emission of charged pairs from the Dirac sea under background electromagnetic fields
has been an interesting topic of constant research since the seminal works by Sauter, Heisenberg-Euler, and Schwinger
\cite{sauter,heisenberg-euler,schwinger}. A strong static electric field can lower the Fermi energy of negative energy, charged particles of the Dirac sea
and open a channel for them to tunnel quantum mechanically into unoccupied, positive energy states and thus create pairs of charged particles as particle-hole pairs. The electromagnetic interaction of the intense background field with the virtual pairs in the Dirac sea makes the vacuum polarized and thereby the Maxwell action replaced by quantum electrodynamics (QED) action \cite{weisskopf,heisenberg-euler,schwinger}.
The pair production and vacuum polarization is a nonperturbative quantum effect that cannot be obtained by summing a finite number
of Feynmann diagrams, and has attracted attention in various areas, such as QED and black hole physics (for review and references, see \cite{dittrich-reuter,dewitt03,RVX}).

The energy consideration puts a constraint on the efficiency of pair-production mechanism in a constant electric field.
A pair of mass $m$ and charge $q$ can be separated over the Compton wavelength $\lambda_C = \hbar c /m$
and be embodied as a real one before being reannihilated
when the electrostatic potential energy between the pair provided is equal to or greater than the rest mass energy $mc^2$.
Thus the critical strength for pair production is $E_C = m^2 c^3/q \hbar$ and $E_C = 1.3 \times 10^{16} {\rm V/cm}$ and $B_C = E_C/c = 4.4 \times 10^{13} {\rm G}$ for electron-positron pair. The collision of heavy atoms may form superatoms
(superheavy atoms) temporarily, whose low energy states can overlap with the Fermi energy of the Dirac sea, and electron-positron pairs can form and positrons can be emitted, leaving a charged vacuum surrounding the nucleus charge \cite{MRG72a,MRG72b} (for review and references, see also \cite{reinhardt-greiner}).
However, magnetars abound in the universe, whose magnetic fields are two order of magnitude larger than the critical strength \cite{harding-lai}.

Charged Reissner-Nordstr\"{o}m black holes may also generate supercritical electric fields and the dyadosphere of charged plasma created by Schwinger mechanism has been proposed as a possible source for gamma rays bursts \cite{PRX,RSWX} (for review and references, see  \cite{RBCFVX,RVX}). The physical process for charged, astrophysical black holes has been challenged since the huge ratio between the gravitational attraction and the electrostatic repulsion prevents any accretion of charges beyond the critical strength \cite{page06}.
Still charged black holes have attracted theoretical interest owing to Hawking's thermal radiation from black holes, and
Schwinger mechanism from charged black holes has also been studied \cite{zaumen,carter,damour-ruffini,gibbons,page77,hiscock-weems,khriplovich,gabriel,kim-page05,CKLSW}. The particle emission from charged black holes consists of the thermal radiation and Schwinger emission of charged particles of the same kind of charge as the black holes. Extremal and near-extremal charged black holes have the zero and small Hawking temperature, so they emit no thermal or exponentially small thermal radiation while the emission is dominated by Schwinger mechanism. Schwinger mechanism of near-extremal charged black holes is the same as that of a constant electric field in ${\rm AdS}_2 \times S^2$ \cite{CKLSW}.

In this paper we shall study Schwinger mechanism by supercritical point charges, which is modeled by a Coulomb potential with large charge $Q$.
This is a toy model for superatoms and charged black holes in that the size of nuclei and the black hole geometry will not be considered. However,
at the asymptotically large distances from the center the electric field behaves Coulomb-like for a spherical charge distribution and
the wave functions may be approximated by those for the Coulomb field. At the 8th Italian-Korean symposium,
with Don N. Page the author presented Schwinger mechanism from a charged Reissner-Nordstr\"{o}m black hole via the instanton action method \cite{kim-page05} (see also Ref. \cite{khriplovich}). The field equation for charged particles in a static electric field in the Minkowski spacetime or a charged black hole geometry describes a tunneling problem and the proper boundary condition yields the instanton action for the tunneling probability for pair production \cite{kim-page02,kim-page06,kim-page07}. The worldline instantons may be used for pair production \cite{dunne-schubert05,DWGS06}.
The main purpose of this paper is to use the wave functions in the supercritical Coulomb field, find the Jost functions and coefficients and therefrom the Bogoliubov transformation. The mean number of the charged bosons with the same charge as the superatom and charged black hole is exactly found. Physical implications of the emission of charged bosons are discussed.

The organization of this paper is as follows. In Sec. II, the instability of the Dirac vacuum is discussed in a supercritical Coulomb field. In Sec. III, the spontaneous emission of charged bosons from the superatom is exactly formulated. In Sec. IV, the spontaneous emission of charged bosons
from a charged black hole is discussed.

\section{Instability of Supercritical Coulomb Field} \label{sec 2}

The Coulomb potential and field of a point-like charge
\begin{eqnarray}
A_0 (r) = \frac{Q}{r}, \quad E(r) = \frac{Q}{r^2}, \label{coulomb}
\end{eqnarray}
can have a supercritical field at the Compton distance $\lambda_C =1/m$ for a particle with mass $m$
and charge $q$ provided that (in unit of $\hbar = c=1$)
\begin{eqnarray}
qQ m^2 \geq m^2. \label{sup E}
\end{eqnarray}
In the case of a superatom (superheavy atom) with charge $Q = Ze$, the condition for the supercritical field is \cite{reinhardt-greiner}
\begin{eqnarray}
Z \alpha \geq 1, \label{sup at}
\end{eqnarray}
where $\alpha = e^2$ is the fine structure constant. Then, the Coulomb field becomes unstable against the spontaneous emission of pairs
of charge $q$ or $e$, known as Schwinger mechanism, and the Dirac vacuum becomes polarized by the Coulomb field. Quantum phenomena such
as the pair production and vacuum polarization in the supercritical Coulomb field also require the one-loop QED action.
In this paper we focus only on the Schwinger mechanism.

In scalar QED, a spinless, charged particle with mass $m$ and charge $e$ satisfies the Klein-Gordon equation in the Coulomb potential with $Q=Ze$
\begin{eqnarray}
\Bigl[- \Bigl(i \frac{\partial}{\partial t} - \frac{Z\alpha}{r} \Bigr)^2 + \frac{1}{r^2} \frac{\partial}{\partial r}
\Bigl(r^2 \frac{\partial}{\partial r} \Bigr) - \frac{{\bf L}^2}{r^2} - m^2 \Bigr] \Phi = 0.
\end{eqnarray}
Expanding the wave function by spherical harmonics
\begin{eqnarray}
\Phi (r, \theta, \varphi) = e^{- i \omega t} Y_{lm} (\theta, \varphi) \frac{\Psi_l (r)}{r}, \label{har sol}
\end{eqnarray}
the radial equation
\begin{eqnarray}
\Bigl[\frac{d^2}{dr^2}
 - \Bigl(\omega - \frac{Z \alpha}{r} \Bigr)^2 + m^2 + \frac{l(l+1)}{r^2} \Bigr] \Psi_l (r) = 0
\end{eqnarray}
has the solution \cite{kim-page05}
\begin{eqnarray}
\Psi_l (r) = A r^{\bar{l} +1} e^{i \sqrt{\bar{\epsilon}}r } M (\bar{l} + 1 + i \bar{\lambda}, 2 \bar{l} +2, -2 i \sqrt{\bar{\epsilon}} r), \label{KG sol}
\end{eqnarray}
where $A$ is a normalization constant and $M$ denotes the confluent hypergeometric function and
\begin{eqnarray}
\bar{l} = - \frac{1}{2} + i C, \quad
\bar{\lambda} = \frac{Z \alpha}{\sqrt{\bar{\epsilon}}/\omega},
\end{eqnarray}
where
\begin{eqnarray}
C = Z \alpha \sqrt{1 - \Bigr( \frac{l+ \frac{1}{2}}{Z\alpha} \Bigr)^2}, \quad
\bar{\epsilon} = \omega^2 - m^2.
\end{eqnarray}
The solutions for spin-1/2 fermions may be found in Ref. \cite{MRG73}

\begin{figure}[t]
{\includegraphics[width=0.65\linewidth,height=0.20\textheight]{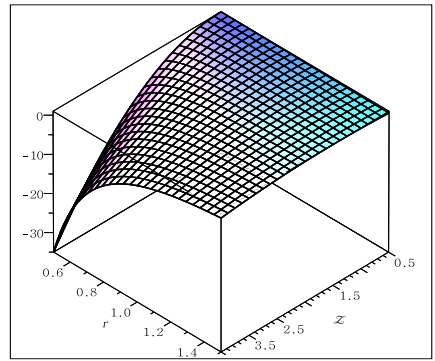} }\hfill
\caption{The effective potential $V (r) = - (\omega - Z\alpha/r)^2 + m^2 + l(l+1)/r^2$ for $l = 0$ and $\omega = 1$ is plotted against ${\cal Z}$ and $r$ in the Compton unit ($m = 1/\lambda_C = 1$) in the range of
$1/2 \leq {\cal Z} \leq 7/2$ and $ 1/2 \leq r \leq 3/2$.} \label{fig-pot1}
\end{figure}

\begin{figure}[t]
{\includegraphics[width=0.65\linewidth,height=0.20\textheight]{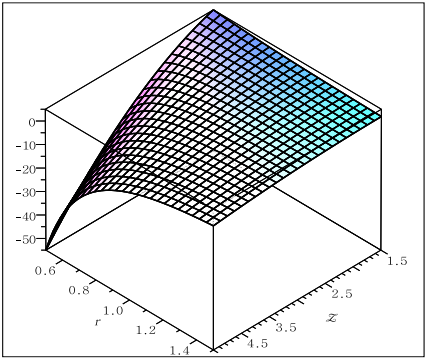} }\hfill
\caption{The effective potential $V (r) = - (\omega - Z\alpha/r)^2 + m^2 + l(l+1)/r^2$ for $l = 1$ and $\omega = 1$ is plotted against ${\cal Z}$ and $r$ in the Compton unit ($m = 1/\lambda_C = 1$) in the range of
$3/2 \leq {\cal Z} \leq 9/2$ and $ 1/2 \leq r \leq 3/2$.} \label{fig-pot2}
\end{figure}

Note that when $Z\alpha < 1/2$ such that $iC < 0$, Eq. (\ref{KG sol}) has a set of bounded states with discrete energy
\begin{eqnarray}
\omega_{nl} = m \Biggl[ 1+  \Bigl(\frac{Z\alpha}{n - (l+ \frac{1}{2}) - iC} \Bigr)^2\Biggr]^{1/2}.
\end{eqnarray}
In the supercritical field (\ref{sup at}), the charged boson $e$ has complex energies corresponding
to the decaying modes (\ref{har sol}), and the Dirac vacuum becomes unstable against the emission of charged bosons
of the same kind of the superatom. The breakdown of the Dirac vacuum may be understood from the effective potential, in which
the bounded states go down below the Dirac sea closer to the nucleus in Compton distances, as shown in Figs. \ref{fig-pot1} and \ref{fig-pot2}.
Thus, charged bosons from the Dirac sea tunnel to unoccupied bounded states forming the charged vacuum
while antibosons with the same kind of charge as the superatom are electrically repelled to infinity, discharging of the superatom \cite{MRG72a,MRG72b}.

\section{Spontaneous Emission of Bosons from Coulomb Potentials}

In the supercritical Coulomb field, the wave functions (\ref{KG sol}) cannot be normalized as those in a constant electric field.
In the latter case one may use the incoming and outgoing wave functions to describe the tunneling problem for pair production
\cite{kim-page05,kim-page02,kim-page06,kim-page07}.
In this paper we employ the quantum field theory for tunneling phenomenon formulated in terms of the Jost functions
for the incoming and outgoing wave functions \cite{kim12}.

For that purpose, using the asymptotic form for the confluent hypergeometric function in the Riemann sheet $-3 \pi /2 \leq z \leq - \pi/2$
\begin{eqnarray}
\frac{M(a,b,z)}{\Gamma (b)} = \frac{e^{-i \pi a}}{\Gamma(b-a)} z^{-a} + \frac{e^{z}}{\Gamma (a)} z^{a-b},
\end{eqnarray}
the solution (\ref{KG sol}) has the asymptotic form at $ r \gg 1/\sqrt{\bar{\epsilon}}$
\begin{eqnarray}
\Psi_l (r) &=& A r^{\bar{l} +1} e^{i \sqrt{\bar{\epsilon}} r} (-2 i \sqrt{\bar{\epsilon}} r)^{- (\bar{l} + 1)} \nonumber\\&& \times
\Bigl[\frac{\Gamma(2 \bar{l} + 2)}{\Gamma(\bar{l} + 1 - i \bar{\lambda})} e^{-i \pi (\bar{l} + 1 + i \bar{\lambda}) }
(-2 i \sqrt{\bar{\epsilon}} r)^{- i \bar{\lambda}} \nonumber\\
&& + \frac{\Gamma(2 \bar{l} + 2)}{\Gamma(\bar{l} + 1 + i \bar{\lambda})}
(-2 i \sqrt{\bar{\epsilon}} r)^{ i \bar{\lambda}} e^{-2 i \sqrt{\bar{\epsilon}} r}
 \Bigr]. \label{asym sol}
\end{eqnarray}
The first and the second terms in the square bracket describe the outgoing and the incoming waves, respectively.
In Ref. \cite{kim12} the Jost functions
\begin{eqnarray}
g(z, K) &=& {\cal C}_1 (K) f(z, -K) + {\cal C}_2 (K) f(z, K), \nonumber\\
g(z, - K) &=& {\cal C} (-K) f(z, K) + {\cal C}_2 (- K) f(z, - K),
\end{eqnarray}
relate the incoming wave $g(z, K)$ and the outgoing wave $g(z, -K)$ at $z = - \infty$ with
another incoming wave $f(z, - K)$ and outgoing wave function $f(z, K)$ at $z =  \infty$.
Thus, the Jost coefficients are
\begin{eqnarray}
{\cal C}_1 (-K) &=& \frac{\Gamma(2 \bar{l} + 2)}{\Gamma(\bar{l} + 1 - i \bar{\lambda})} e^{-i \pi (\bar{l} + 1 + i \bar{\lambda}) } e^{i \frac{\pi}{2} (\bar{l} + 1 + i \bar{\lambda})}, \nonumber\\
{\cal C}_2 (-K) &=& \frac{\Gamma(2 \bar{l} + 2)}{\Gamma(\bar{l} + 1 + i \bar{\lambda})} e^{i \frac{\pi}{2} (\bar{l} + 1 - i \bar{\lambda})},
\end{eqnarray}
where $K$ denotes quantum numbers $K = (\omega, l, m)$.
Then the Bogoliubov coefficients
\begin{eqnarray}
\mu_K = \frac{{\cal C}_1 (-K)}{{\cal C}_2^* (-K)}, \quad \nu_K = \frac{1}{{\cal C}_2^* (-K)},
\end{eqnarray}
are found to be
\begin{eqnarray}
\mu_K &=& \frac{\Gamma(2 \bar{l} + 2)}{\Gamma(2 \bar{l}^* + 2)} \frac{\Gamma(\bar{l}^* + 1 - i \bar{\lambda})}{\Gamma(\bar{l} + 1 - i \bar{\lambda})}
e^{- i \frac{\pi}{2} (\bar{l} - \bar{l}^*)}, \nonumber\\
\nu_K &=& \frac{\Gamma(\bar{l}^* + 1 - i \bar{\lambda})}{\Gamma(2 \bar{l}^* + 2)} e^{ i \frac{\pi}{2} (\bar{l}^* +1 + i \bar{\lambda}^* )}.
\end{eqnarray}
Finally, the vacuum persistence and pair-production rate is given by
\begin{eqnarray}
\vert \mu_K \vert^2 &=& \frac{\cosh \pi(\bar{\lambda} - C)}{\cosh \pi(\bar{\lambda} + C)} e^{2 \pi C}, \nonumber\\
\vert \nu_K \vert^2 &=& \frac{\sinh (2 \pi C)}{\cosh \pi(\bar{\lambda}+C )} e^{- \pi (\bar{\lambda}- C)},
\end{eqnarray}
and they satisfy the Bogoliubov relation $ \vert \mu_K \vert^2 - \vert \nu_K \vert^2 =1$.
Noting that $\bar{\lambda} > C$, the mean number of pairs can be written as
\begin{eqnarray}
\bar{N} (\omega, l) = \vert \nu_K \vert^2 = e^{- 2 \pi (\bar{\lambda}- C)} \frac{1+ e^{-4 \pi C}}{1+ e^{-2 \pi(\bar{\lambda}+C )}}.
\end{eqnarray}
The mean number is approximately $\bar{N} \approx e^{- 2 \pi (\bar{\lambda}- C)}$ for low angular momenta $l \ll Z \alpha$ but
$\bar{N} \approx 2 e^{- 2 \pi \bar{\lambda}}$ for high angular momenta $l \approx Z \alpha$.
The total mean number of pairs per unit time is the sum of mean number over each energy and angular momentum
\begin{eqnarray}
\bar{N} (\omega, l) = \sum_{l = 0}^{\infty} (2l+1) \int_{m}^{\infty} \frac{ d \omega}{2 \pi}  \bar{N} (\omega, l).
\end{eqnarray}

\begin{figure}[t]
{\includegraphics[width=0.65\linewidth,height=0.20\textheight]{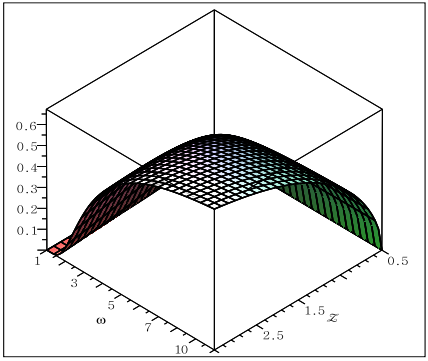} }\hfill
\caption{The mean number of pairs for $l = 0$ is plotted against ${\cal Z} = Z \alpha$ and $\omega$ in unit of $m =1$ in the range of
$1/2 \leq {\cal Z} \leq 7/2$ and $ 1 \leq \omega \leq 10$.} \label{fig-num1}
\end{figure}

\begin{figure}[t]
{\includegraphics[width=0.65\linewidth,height=0.20\textheight]{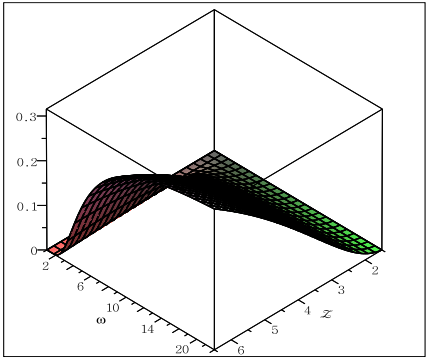} }\hfill
\caption{The mean number of pairs for $l = 1$ is plotted against ${\cal Z} = Z \alpha$ and $\omega$ in the range of
$3/2 \leq {\cal Z} \leq 13/2$ and $ 1 \leq \omega \leq 20$.} \label{fig-num2}
\end{figure}

\begin{figure}[t]
{\includegraphics[width=0.65\linewidth,height=0.20\textheight]{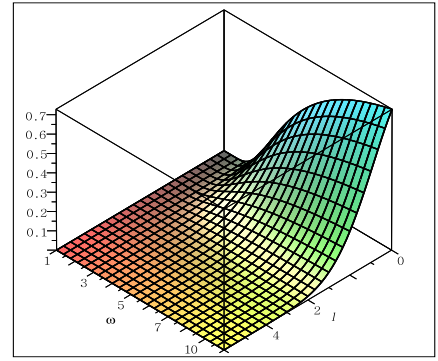} }\hfill
\caption{The mean number of pairs for ${\cal Z} = Z \alpha = 11/5$ is plotted against $l$ and $\omega$ in the range of
$0 \leq l \leq 4$ and $ 1 \leq \omega \leq 10$.} \label{fig-an-ch}
\end{figure}

The emission rate of charged bosons of the same kind of charge as
the superatom increases as the charge $Z\alpha$ and the energy $\omega$, as shown in Figs. \ref{fig-num1} and \ref{fig-num2}.
It is suppressed for large angular momenta due to the centrifugal repulsion but is also
favored for large charge and energy for a given angular momentum, as shown in Fig. \ref{fig-an-ch}.
Note that the emission of charged bosons has a threshold for the charge of the superatom
\begin{eqnarray}
Z \alpha = eQ \geq l + \frac{1}{2}.
\end{eqnarray}

\section{Spontaneous Emission from Charged Black Holes} \label{black hole}

In general relativity a charge with a large mass
can form a black hole. A point charge $Q$ with mass $M$ with the Coulomb potential (\ref{coulomb})
has the charged Reissner-Nordstr\"{o}m black hole with the metric
(in the geometric unit $\hbar = c = G=1$)
\begin{eqnarray}
ds^2 = - \Delta (r) dt^2 + \frac{dr^2}{\Delta (r)} + r^2 d \Omega_2^2,
\end{eqnarray}
where in terms of the outer and inner horizons $r_{\pm} = M \pm \sqrt{M^2 - Q^2}$
\begin{eqnarray}
\Delta (r) = \Bigl(1 - \frac{r_{+}}{r} \Bigr) \Bigl(1 - \frac{r_{-}}{r} \Bigr).
\end{eqnarray}
The spherical harmonics of a charged boson obeys the radial equation \cite{kim-page05}
\begin{eqnarray}
\Bigl[ \frac{d^2}{dr_*^2} - V_l(r)  \Bigr] \Psi_l (r) = 0,
\end{eqnarray}
where the tortoise coordinate is
\begin{eqnarray}
r_* = r + \frac{r_+^2}{r_+ - r_-} \ln(r - r_+) - \frac{r_-^2}{r_+ - r_-} \ln(r - r_-),
\end{eqnarray}
and the effective potential is
\begin{eqnarray}
V_l (r) &=& \Delta (r) \Bigl( m^2 + \frac{l(l+1)}{r^2} + \frac{2M}{r^3} - \frac{2Q^2}{r^4} \Bigr)
\nonumber\\ && - \Bigl( \omega - \frac{qQ}{r} \Bigr)^2.
\end{eqnarray}
The asymptotic form of the effective potential is approximately given by that of the Coulomb potential
\begin{eqnarray}
V_l (r) \approx  m^2 + \frac{l'(l'+1)}{r^2}  - \Bigl( \omega - \frac{qQ'}{r} \Bigr)^2
\end{eqnarray}
with the modified charge and angular momenta
\begin{eqnarray}
Q' &=& Q + \frac{m^2 M}{q \omega}, \nonumber\\
l' &=& \sqrt{\Bigl(l+ \frac{1}{2} \Bigr)^2 + \frac{m^2 M}{\omega} \Bigl(2qQ + \frac{m^2 M}{\omega}
\Bigr)} - \frac{1}{2}.
\end{eqnarray}
The emission rate of the same kind of charge as the black hole is
\begin{eqnarray}
N (\omega, l) \approx \frac{\sinh 2 \pi C'}{\cosh \pi( C' + \bar{\lambda'} )} e^{\pi (C' - \bar{\lambda'})}. \label{bh pair}
\end{eqnarray}

A few remarks are in order. First, note that
\begin{eqnarray}
C' = \sqrt{(qQ')^2 - \Bigl(l' + \frac{1}{2} \Bigr)^2} = C
\end{eqnarray}
is an invariant. Second, the emission rate (\ref{bh pair}) of charged bosons has the same form as Eqs. (34) and (40) of Ref. \cite{CKLSW} for
the extremal black hole with the correspondence
\begin{eqnarray}
 \bar{\lambda'} &\Leftrightarrow& a = qQ, \nonumber\\
 C' &\Leftrightarrow& b = qQ \sqrt{ 1 - \Bigl(\frac{l+\frac{1}{2}}{qQ} \Bigr)^2 - \frac{m^2}{q^2}}.
\end{eqnarray}

\section{Conclusion} \label{conclusion}

We have studied the spontaneous emission of charged bosons from the supercritical Coulomb field of a point charged superatom with $Z \alpha > 1$.
The supercritical Coulomb field of a point charge may provide approximately an analytical toy model for finite superatoms formed from collision of heavy atoms and charged Reissner-Nordstr\"{o}m black holes. The energy of the bounded states in an undercritical Coulomb field becomes complex when $Z \alpha$ is analytically continued over the critical value $Z \alpha = l + 1/2$ for angular momentum $l$ and the bounded states decay and
their wave functions cannot be normalized. The bounded states whose energy is lower than the Fermi
energy of the Dirac sea can be occupied by the charged bosons penetrated from the Dirac sea and form charged vacuum \cite{MRG72a,MRG72b}.

In this paper we have employed the Jost functions for the wave functions for the supercritical Coulomb field,
found the Bogoliubov transformation from the tunneling boundary condition and obtained the exact formulation of the pair-production rate.
A superatom discharges by emitting bosons of the same kind of charge as the atom. The mean number of emitted bosons increases as the charge of the superatom or black hole and the energy of the bosons increases but decreases as the angular momentum increases.
The supercritical Coulomb potential of a point charge may approximately prescribe the electric potential of
a charged Reissner-Nordstr\"{o}m black hole with a modified charge and angular momenta in the asymptotic region
and results in the emission rate of bosons from the black hole. Not pursued in this paper is the vacuum polarization
due to the supercritical Coulomb field. The Bogoliubov coefficients in this paper may be used for finding
the one-loop effective action in the in-out formalism and thereby the vacuum polarization and the pair production,
which will be addressed in a future publication.

\acknowledgments
The author thanks Remo Ruffini and She-Sheng Xue for useful discussions at the 13th Italian-Korean Symposium on Relativistic
Astrophysics at Ewha Womans University on July 15-19, 2013.
He also thanks Christian Schubert for the warm hospitality at Instituto de F\'{i}sica y Matem\'{a}ticas, Universidad Michoacana de San Nicol\'{a}s de Hidalgo, where this paper was initiated, and Chang Hee Nam for the warm hospitality at the Center for Relativistic Laser Science (CoReLS), Institute for Basic Science (IBS), where this paper was completed. He thanks Wei-Tou Ni for reading the manuscript.
This work was supported in part by Basic Science Research Program through the National Research Foundation of Korea (NRF) funded by the Ministry of Education (NRF-2012R1A1B3002852).

\end{document}